\documentclass[prb,aps,twocolumn,superscriptaddress,showpacs]{revtex4}
\usepackage{graphicx}
\usepackage{bm}
\usepackage{amsmath}
\renewcommand{\bbox}[1]{\bm{#1}}

\begin{document}
\title{Exciton-LO-phonon dynamics in InAs/GaAs quantum dots:
effects of zone-edge phonon damping}

\author{Pawe{\l} Machnikowski}
\affiliation{Institute of Physics, Wroc{\l}aw University of
Technology, 50-370 Wroc{\l}aw, Poland}
\affiliation{Institute f\"ur Festk\"orpertheorie, Universit\"at M\"unster,
48149 M\"unster, Germany}
\author{Lucjan Jacak}
\affiliation{Institute of Physics, Wroc{\l}aw University of
Technology,
50-370 Wroc{\l}aw, Poland}

\begin{abstract}
The dynamics of an exciton-LO-phonon system after an ultrafast
optical excitation in an InAs/GaAs quantum dot is studied
theoretically. Influence of anharmonic phonon damping and its
interplay with the phonon dispersion is analyzed. The signatures
of the zone-edge decay process in the absorption spectrum and time
evolution are highlighted, providing a possible way of
experimental investigation on phonon anharmonicity effects.
\end{abstract}


\maketitle

\section{Introduction}

Unlike natural atoms, semiconductor quantum dots (QDs) always form
part of a macroscopic crystal. The interaction with the
quasi-continuum of lattice degrees of freedom (phonons)
constitutes an inherent feature of these nanometer-size systems
and cannot be neglected in any realistic modeling of QD
properties, especially when the coherence of confined carriers is
of importance. The understanding of the decisive role played by
the carrier coupling to lattice modes has increased recently due
to both experimental and theoretical study, including the effects of the 
carrier-phonon interaction in a system driven optically on ultrafast 
time scales
\cite{borri,krummheuer02,vagov,jacak03b}.
The phonon-induced decoherence
seems to be crucial for any quantum information processing
application and for any nanotechnological device relying on
quantum coherence of confined carriers
\cite{alicki04,machnikowski04b}.

One of the issues of interest is the evolution of the
carrier-lattice system after ultrafast excitation of carriers.
Experiments performed on bulk and quantum well systems
\cite{cho90,banyai95,wehner98a,furst97,wehner98b} show strongly
non-Markovian behavior of carrier relaxation on 100 fs timescale
due to pertinent system memory, yielding access to the quantum
kinetic regime of the system dynamics where fundamental ideas of
quantum mechanics may be tested.

For
systems confined in all 3 dimensions,
where purely longitudinal optical (LO) phonon
relaxation is forbidden by
energy conservation, coherent LO phonon dynamics may still be induced
by lattice
relaxation after non-adiabatic change of carrier distribution (dressing
process \cite{jacak03b}).
In this process, an ultrafast excitation of confined carriers results in
the lattice state which no longer corresponds to the energy minimum and
therefore is unstable. Before the excess energy may be transferred outside
the QD area (by radiating out phonons), a transient dynamics takes place,
manifested by the oscillations (beats, overdamped for acoustical phonons)
in the optical response of the system\cite{borri}.

Femtosecond time-resolved experiments showing the joint LO-phonon
evolution (beats) in such quasi-0-dimensional systems were
performed only on semiconductor nanocrystals
\cite{roussignol89,schoenlein93,mittleman94}, where confined and
surface modes come into play, leading to different properties of
the carrier-phonon coupling (e.g. size dependence
\cite{jacak03a}). For self-assembled semiconductor QDs, only 1 ps
timescale dynamics was reported \cite{borri}. However,
some theoretical models, developed for the description of
nonlinear optical response in bulk or 2D systems
\cite{steinbach99,axt99}, may be successfully applied to the
confined (0D) case with discrete carrier spectrum. Within such
independent boson models it was possible to describe the
linear\cite{krummheuer02,besombes01,pazy02} and
nonlinear\cite{vagov} response in the limit of infinitely short
pulses and to account quantitatively\cite{vagov} for the
short-time behavior of optical polarization observed in the
time-resolved spectroscopy experiment \cite{borri}.
Although the observed effects may be attributed only to acoustic
phonons \cite{vagov}, the theory predicts also LO phonon
effects for sufficiently short pulses
\cite{jacak03a,krummheuer02}.

Due to the weak dispersion of LO modes effectively coupled to the
confined carriers, the dephasing of LO phonon beats is extremely
slow (100 ps timescale) \cite{jacak03b}. On the other hand,
anharmonic damping reduces the lifetime of a zone-center LO phonon
in GaAs to 9.2 ps at low temperatures \cite{vallee91}.
Hence, including the anharmonicity is essential for a correct
description of the confined carrier-LO-phonon dynamics. The actual
nature of anharmonic phonon-phonon interactions is, unfortunately,
still unclear. Experimental results for the bulk GaAs
\cite{vallee91} suggest dominant role of the decay
channel to zone-edge LO and transversal acoustical (TA) phonons,
although another interpretation of the same data proposes a major
contribution of the decay into two longitudinal acoustical (LA)
phonons \cite{barman02}. The latter channel has been also studied
theoretically both for bulk \cite{bhatt94} and confined
\cite{li98} LO phonons. Quite unexpectedly, phonon dispersion
measurements \cite{strauch90} suggest that the zone-edge process
may be forbidden by energy and momentum conservation. Since the
anharmonic effects seem to be essential e.g. for correct modeling
of relaxation phenomena
\cite{verzelen00,verzelen02b,jacak02a,jacak03a}, this ambiguity is
a real problem.

In the present paper we describe theoretically the quantum evolution of LO
phonons damped by anharmonicity after an ultrafast carrier
(exciton) excitation. We focus on the effects of the zone-edge
decay process and its possible signatures in the spectral line
shapes and temporal dynamics. We single out various regimes both
for single mode evolution and for the collective dynamics of all
the coupled phonons. For each of these cases we calculate the
linear optical absorption for frequencies corresponding to LO
phonon peaks as well as the evolution of
the reduced density matrix after an ultrafast excitation. Both
characteristics (in frequency and time domain) are calculated
within a single formalism, based on the Green function technique,
restricted to single-phonon effects (which is a reasonable
approximation in the weak coupling case at low temperatures) but
incorporating the contribution from optically inactive exciton states 
excited by
nonadiabaticity of LO phonon dynamics \cite{fomin00}. We show that
the zone-edge decay of LO phonons manifests itself clearly in the
spectral properties of the system for certain energy relations
between the zone-center and zone-edge phonons. Although these
energy relations are a material feature under fixed conditions, in
view of the existing uncertainty concerning the phonon
anharmonicity channels it seems interesting to study all possible
scenarios. Moreover, even if the zone-edge process turns out to be
forbidden at normal pressure, it is likely to appear at high
pressures. Indeed, the measured pressure (volume) dependence of
phonon frequencies\cite{trommer80} shows that the energy of the
zone-center LO phonon increases with growing pressure, while the
total energy of the zone-edge LO and TA phonon pair slowly
decreases. The cross-over pressure and the phonon replica
line shapes close to the cross-over would yield information both on
the phonon energy relations and on the anharmonic coupling
strength.

The paper is organized as follows: The section \ref{sec:model} introduces
the model and discusses the formal quantities conveniently characterizing
the system dynamics. Technical derivation is contained in the section
\ref{sec:green}. The physical discussion of the result is given in the
section \ref{sec:results}. The section \ref{sec:conclusions} concludes
the paper with final remarks.

\section{The model and the formalism}
\label{sec:model}

We consider the Hamiltonian describing a single exciton
interacting with LO phonons,
\begin{equation}
H=H_{\mathrm{X}} +H_{\mathrm{ph}} + H_{\mathrm{int}}.
\label{ham}
\end{equation}
The first part is the exciton Hamiltonian,
\begin{displaymath}
H_{\mathrm{X}}=  \sum_{n}\epsilon_{n}a^{+}_{n}a_{n},
\end{displaymath}
where $\epsilon_{n}$ is the energy of the $n$th exciton state and
$a_{n},a_{n}^{\dagger}$ are the corresponding annihilation and
creation operators. Our discussion will be simplified by using the
specific structure of the excitonic spectrum in self-assembled
InAs/GaAs systems\cite{jacak03b}. Due to the much larger hole
effective mass, the lowest excited states are formed to a good
approximation by exciting the hole. Therefore, the lowest excited
states of the confined exciton may be modeled by products of the
corresponding harmonic oscillator wave functions
($\sim\exp[-(r_{i}/l_{i})^{2}/2]$; ground state for the electron and
ground or excited state for the hole), with Gaussian widths $l_{i}$
given in Table \ref{tab:param}. The resulting excited two-particle
states are either strictly optically inactive due to parity or almost
inactive because the ground state electron and excited state hole
wave functions are close to orthogonal even in absence of strict
selection rules. The excitation energy typical for such inactive
states is $\sim 15$ meV, consistent with the thermally activated
contribution to the central line broadening measured in experiment
\cite{borri}. The lowest optically active state involves excitation of
both the electron and the hole and its energy is typically much higher
($\sim 70$ meV). Thus, even the LO replicas of the lowest excited
state(s) lie still below the active state. Under totally non-selective
excitation all states contribute to the dynamics and
the resulting evolution would be both difficult to describe
theoretically and hard to interpret experimentally. Therefore, we
assume some degree of spectral selectivity due to finite pulse length,
sufficient to avoid the excitation of the optically active excited
state, and include in our description only the lowest, optically
inactive excited state.

The phonon Hamiltonian $H_{\mathrm{ph}}$ describes the LO and TA
phonon branches coupled by anharmonicity\cite{vallee91},
\begin{eqnarray*}
H_{\mathrm{ph}} & = & \sum_{{\bbox{k}}}\Omega_{\bbox{k}}
b_{\bbox{k}}^{\dagger}b_{\bbox{k}}
+\sum_{{\bbox{k}}}\omega_{\bbox{k}}
c_{\bbox{k}}^{\dagger}c_{\bbox{k}} \\
& & +\sum_{\bbox{k},\bbox{k'}}W(\bbox{k},\bbox{k}')
b_{\bbox{k}}b^{\dagger}_{\bbox{k}-\bbox{k}'}c^{\dagger}_{\bbox{k}'}
+\mathrm{H.c.},
\end{eqnarray*}
where $b_{\bbox{k}},b_{\bbox{k}}^{\dagger}$ and
$c_{\bbox{k}},c_{\bbox{k}}^{\dagger}$ refer to LO and TA phonons
with momentum $\bbox{k}$, respectively, and $\Omega_{\bbox{k}}$,
$\omega_{\bbox{k}}$ are the corresponding energies. To simplify
the notation, we put    $\hbar=1$, $k_{\mathrm{B}}=1$. The
anharmonic term describes a decay of LO phonon into a TA phonon
and another LO phonon, which may be allowed by energy and momentum
conservation only when  the final states lie in the vicinity of
the $L$ point of the Brillouin
zone\cite{vallee91,strauch90}. The interaction term
accounts for the Fr\"ohlich interaction between confined charges
and the lattice polarization induced by the LO deformation,
\begin{equation}
H_{\mathrm{int}}=
\frac{1}{\sqrt{N}}\sum_{{\bbox{k}},n,n'}
F_{n,n'}({\bbox k}) a_{n}^{+}a_{n'}
\left( b_{{\bbox k}}+b_{{-\bbox k}}^{\dagger} \right),
\end{equation}
where the coupling constants for the confined carriers are given
by
\begin{eqnarray}\label{fo}
F_{n,n'}( \bbox k) & = & -\frac{e}{k}
\sqrt{\frac{\hbar\Omega_{0}}{2v\varepsilon_{0}\tilde{\varepsilon}}}
\int d^{3}{\bbox{r}_{\mathrm{e}}}
d^{3}{\bbox{r}_{\mathrm{h}}} \\
& &  \times\Psi^{*}_{n} (\bbox{r}_{\mathrm{e}},
\bbox{r}_{\mathrm{h}})
\left(e^{i\bbox{k}\cdot\bbox{r}_{\mathrm{e}}} -
e^{i\bbox{k}\cdot\bbox{r}_{\mathrm{h}}}\right)
\Psi_{n'}(\bbox{r}_{\mathrm{e}}, \bbox{r}_{\mathrm{h}}). \nonumber
\end{eqnarray}
where $\bbox{r}_{\mathrm{e}}, \bbox{r}_{\mathrm{h}}$ denote
the coordinates of the electron and hole, respectively, 
$\epsilon_{0}$ is the vacuum dielectric constant,
and the other elements of the notation are described in
Table \ref{tab:param}, along with
values (corresponding to InAs/GaAs system) used in the calculations.

\begin{table}[tb]
\begin{tabular}{lll}
\hline
Effective dielectric constant\cite{adachi85} &
$\tilde{\varepsilon}$ & 62.6 \\
Optical phonon energy\cite{strauch90} & $\hbar\Omega_{0}$ & $36$
meV \\
Unit crystal cell volume & $v$ & 0.044 nm$^{3}$ \\
$\Gamma$ point LO dispersion parameter\cite{strauch90} &
$\beta_{\mathrm{LO},\Gamma}$ & -0.04 meV$\cdot$nm$^{2}$ \\
$L$ point LO dispersion parameters\cite{strauch90}: & & \\
\hspace{0.7em}parallel to the zone boundary &
$\beta_{\mathrm{LO},||}$
   & 0.17 meV$\cdot$nm$^{2}$ \\
\hspace{0.7em}perpendicular to the z. b. &
$\beta_{\mathrm{LO},\perp}$
   & 0.35 meV$\cdot$nm$^{2}$ \\
$L$ point TA dispersion parameters\cite{strauch90}: & & \\
\hspace{0.7em}parallel to the zone boundary &
$\beta_{\mathrm{TA},||}$
   & 0.16 meV$\cdot$nm$^{2}$ \\
\hspace{0.7em}perpendicular to the z. b. &
$\beta_{\mathrm{TA},\perp}$
   & -0.04 meV$\cdot$nm$^{2}$ \\
Zone-center phonon lifetime\cite{vallee91} &
$\tau_{\mathrm{LO}}$
   & 9.2 ps \\
Wave function widths: & & \\
\hspace{0.7em} electron, in-plane & $l_{\mathrm{e},\perp}$ & 4.0 nm  \\
\hspace{0.7em} hole, in-plane & $l_{\mathrm{h},\perp}$ & 3.3 nm  \\
\hspace{0.7em} electron and hole, $z$-direction & $l_{z}$ & 1 nm  \\
\hline
\end{tabular}
\caption{\label{tab:param}The parameters used in the
calculations.}
\end{table}

The system properties may be experimentally studied in the
frequency or time domain. Although usual spectral functions describe 
the equilibrium properties of the system and are relevant only
in the linear response regime, in the limiting case of ultrafast excitation
they are sufficient for the description of the system evolution also in the 
strongly driven case. This is possible due to separation of the time scales 
(external driving vs. phonon response): the laser pulse prepares an initial 
state while the subsequent evolution consists in relaxation to equilibrium. 

To see this more formally, let us note that the electromagnetic wave is
coupled only to carrier degrees of freedom and not to lattice
modes so that all information accessible to optical measurement is given
by the reduced density matrix of the carrier subsystem,
\begin{displaymath}
\varrho(t)
={\mathrm{Tr}}_{\mathrm{L}}[U(t)\rho(0)U^{\dagger}(t)],
\end{displaymath}
where $\rho(0)$ is the initial state of the total system
and the trace is taken over the lattice (phonon) degrees of freedom.
Assuming that the state $n$ is the only relevant
optically active state in the frequency sector of interest,
the optical properties are determined by the non-diagonal element
of the reduced density matrix of the carrier subsystem, proportional to
the optical polarization,
\begin{equation}
\varrho_{n}(t)=\langle {\mathrm{vac}} | \varrho(t) |n\rangle
={\mathrm{Tr}}\left[ \rho(0)a_{n}(t) \right], \label{densmat}
\end{equation}
where $| \mathrm{vac} \rangle$ stands for vacuum (empty dot). 
Let us consider a change in the carrier subsystem performed by the
ultrafast pulse that takes the initial vacuum state
instantaneously into a certain superposition of the vacuum state and the 
single-exciton state $n$, i.e., $|\psi\rangle=\cos\frac{\alpha}{2}
| {\mathrm{vac}} \rangle +\sin\frac{\alpha}{2}a_{n}^{\dagger}|
{\mathrm{vac}} \rangle$, where $\alpha$ is the area of the ultrafast 
driving pulse.
Due to its inertia, the lattice is left
in its initial state, i.e. in thermal equilibrium described by the
density matrix $\rho_{\mathrm{L}}$.  The further dynamics, after
switching off the pulse, is generated by the Hamiltonian
(\ref{ham}), conserving the number of excitons. Inserting the
initial state $\rho(0)=|\psi\rangle\langle\psi |\otimes
\rho_{\mathrm{L}}$ into Eq.~(\ref{densmat}) one gets
\begin{displaymath}
\varrho_{n}=\frac{1}{2}\sin \alpha I_{n}(t),
\end{displaymath}
where the exciton correlation function is defined by
\begin{equation}
I_{n}(t)=\langle a_{n}(t)a_{n}^{\dagger}(0)\rangle
\label{correl}
\end{equation}
($\langle\cdot\rangle$ denotes the thermal average). Thus, the arbitrary 
angle $\alpha$ factors out and the evolution is described universally by a 
single function $I(t)$.

The function (\ref{correl}) corresponds to the ``survival amplitude'', 
i.e., to the overlap between the state of an exciton created
instantaneously at $t=0$ and evolving under the full carrier-phonon 
Hamiltonian until the time $t$ and the unperturbed exciton state. For our
discussion it is essential that this correlation function is 
related to the standard Green functions formalism. Indeed, 
defining the retarded Green function 
\begin{displaymath}
 G_{nn'}(\omega)=\int dt e^{i\omega t}
 \Theta(t)\langle \{a_n(t),a^{\dagger}_{n'}\} \rangle
\end{displaymath}
and the corresponding spectral density
\begin{equation}\label{Aw}
    A_{n}(\omega)=-2{\mathrm{Im}}G_{nn}(\omega),
\end{equation}
one can write \cite{mahan00}
\begin{eqnarray}\label{An}
A_{n}(\omega) & = & \left( 1-e^{-\omega/T} \right)^{-1}
\int_{-\infty}^{\infty}I_{n}(t)e^{i\omega t}dt \\
& \simeq & \int_{-\infty}^{\infty}I_{n}(t)e^{i\omega t}dt.
\nonumber
\end{eqnarray}
The final approximation is justified by the fact that for a
typical semiconductor, the exciton energy is 1--2 orders of
magnitude larger than the thermal energy even at room temperature,
so the Boltzmann factor in this formula may be neglected. The
experimentally measurable absorption is proportional
to $A_{n}(\omega)$.
Hence, both the linear spectrum and the evolution after a (possibly strong)
ultrashort pulse is conveniently described within the Green function
approach.

\section{Exciton Green function}
\label{sec:green}

In order to find the diagonal Green function 
$G_{nn}(\omega)$ we write the 
Dyson equation in the Matsubara formalism,
\begin{displaymath}
G_{nm}(ip_{\nu})=G_{nm}^{(0)}(ip_{\nu})
  +G_{nl}^{(0)}(ip_{\nu})\Sigma_{ll'}(ip_{\nu})G_{l'm}(ip_{\nu}).
\end{displaymath}
In order to approximately invert this system of equations we note that the
diagonal Green functions are of zeroth order in the carrier-phonon
interaction while the non-diagonal ones, as well as the the mass operator 
elements $\Sigma_{nn'}$, are at least of the 
second order. Hence, the system may be inverted to the leading order to 
yield the diagonal exciton Green function 
\begin{displaymath}
G_{n}(ip_{\nu})\equiv G_{nn}(ip_{\nu})
=\frac{1}{ip_{\nu}-\epsilon_{n}-\Sigma_{n}(ip_{\nu})},
\end{displaymath}
where $\Sigma_{n}(ip_{\nu})\equiv \Sigma_{nn}(ip_{\nu})$. 

Throughout the paper, the Greek
indices will be used for numbering Matsubara frequencies, while
the Latin ones for exciton states (including all the relevant
quantum numbers). In the single-exciton approximation, the exciton
may be treated as a fermion, with the corresponding frequencies
$p_{\nu}$. The bosonic frequencies are denoted by $\omega_{\mu}$,
while $\epsilon_{n}$ denotes the bare exciton energy in the
eigenstate $n$.

The corresponding retarded functions are
\begin{eqnarray*}
G_{n}(\omega) & = &
G_{n}(ip_{\nu})|_{ip_{\nu}\rightarrow\omega+i\delta} \\
\Sigma_{n}(\omega) & = &
\Sigma_{n}(ip_{\nu})|_{ip_{\nu}\rightarrow\omega+i\delta}.
\end{eqnarray*}
Writing the retarded mass operator as
$\Sigma_{n}(\omega)=\Delta_{n}(\omega)-i\gamma_{n}(\omega)$, one has
\begin{equation}
G_{n}(\omega)=
\frac{1}{\omega-\epsilon_{n}-\Delta_{n}(\omega)
	+i\gamma_{n}(\omega)+i\delta}.
\label{Green}
\end{equation}
The energy
levels $E_{n}$ for the exciton interacting with LO phonons
(the confined excitonic polaron \cite{verzelen02a,jacak03b}) may be found as
the poles of this function; their real parts are given by the equation
\begin{equation}
E_{n}-\epsilon_{n}-\Delta_{n}(E_{n})=0.
\label{del}
\end{equation}
Let us note that the correction $\Delta_{n}(\omega)$
in the denominator of Eq.~(\ref{Green}) is
important only near $\epsilon_{n}$, where the other term vanishes
(otherwise, it is a higher-order correction).
Therefore, it may be replaced by
\begin{displaymath}
\Delta_{n}(\omega)\approx \Delta_{n}(E_{n})
+(\omega-E_{n})\left( \frac{d\Delta_{n}(\omega)}{d\omega}
\right)_{\omega=E_{n}}
\end{displaymath}
and combined with $\epsilon_{n}$ to give $E_{n}$,
in accordance with Eq.~(\ref{del}).

Thus, one may write
\begin{equation}\label{Gw}
    G_{n}(\omega)
 = Z_{n}^{-1}\frac{1}{\omega-E_{n}+i\gamma_{n}(\omega)},
\end{equation}
where
\begin{equation}
Z_{n}=1-\frac{d\Delta_{n}(\omega)}{d\omega}|_{\omega=E_{n}}.
\label{Z}
\end{equation}

Restricting the discussion to the weak coupling case, we are
interested in the single-phonon approximation, except for
including the lowest-order anharmonicity effects into the LO
phonon propagation. Therefore, we neglect the vertex correction in
the mass operator $\Sigma_{n}(ip_{\nu})$ and replace the exact
phonon Green function by the free one perturbed by the coupling to
TA phonons in the leading order. The mass operator in this
approximation is given by
\begin{eqnarray}\label{mass1}
\lefteqn{\Sigma_{n}(ip_{\nu})=}
\\
& & -\frac{1}{\beta}\frac{1}{N}\sum_{n',\bbox{k}}
\sum_{\mu}|F_{nn'}(\bbox{k})|^{2}G_{n'}(ip_{\nu}-i\omega_{\mu})
D(\bbox{k},i\omega_{\mu}), \nonumber
\end{eqnarray}
and the LO phonon Green function is
\begin{equation}
D(\bbox{k},i\omega_{\mu})=-\frac{2\Omega_{\bbox{k}}}%
{-(i\omega_{\mu})^{2}+\Omega_{\bbox{k}}^{2}
+2\Omega_{\bbox{k}}\Pi(\bbox{k},i\omega_{\mu})}.
\label{phonGreen}
\end{equation}
The polarization operator in
the proposed approximation involves the lowest order anharmonicity
correction
\begin{widetext}
\begin{equation}
\Pi(\bbox{k},i\omega_{\mu})=
-\frac{1}{\beta}\frac{1}{N}\sum_{\bbox{k}',\mu'} |W(\bbox{k},\bbox{k}')|^{2}
D^{(0)}_{\mathrm{LO}}(\bbox{k}-\bbox{k}',i\omega_{\mu}-i\omega_{\mu'})
D^{(0)}_{\mathrm{TA}}(\bbox{k}',i\omega_{\mu'}).
\label{polar1}
\end{equation}
Inserting the free phonon Green functions into Eq.~(\ref{polar1})
and performing the summation
over frequencies one gets
\begin{eqnarray}\label{polar2}
\lefteqn{\Pi(\bbox{k},i\omega_{\mu})=}\\
\nonumber
&& -\frac{2}{N}\sum_{\bbox{k}'}
|W(\bbox{k},\bbox{k}')|^{2} \left[
\frac{(n_{\bbox{k}'}-N_{\bbox{k}-\bbox{k}'})
(\Omega_{\bbox{k}-\bbox{k}'}-\omega_{\bbox{k}'})}
{(i\omega_{\mu})^{2}-(\Omega_{\bbox{k}-\bbox{k}'}-\omega_{\bbox{k}'})^{2}}
+\frac{(n_{\bbox{k}'}+N_{\bbox{k}-\bbox{k}'}+1)
(\Omega_{\bbox{k}-\bbox{k}'}+\omega_{\bbox{k}'})}
{(i\omega_{\mu})^{2}-(\Omega_{\bbox{k}-\bbox{k}'}+\omega_{\bbox{k}'})^{2}}
\right],
\end{eqnarray}
where $N_{\bbox{k}}$ and $n_{\bbox{k}}$ are the Bose distributions
for LO and TA phonons, respectively, with momentum $\bbox{k}$.

The Green functions may be written in the representation
\begin{equation}\label{spectral}
D(\bbox{k},i\omega_{\mu})  =
\int_{-\infty}^{\infty}\frac{d\omega}{2\pi}
\frac{B(\bbox{k},\omega')}{i\omega_{\mu}-\omega'},\;\;\;
G_{n'}(i\omega_{\mu}) =
\int_{-\infty}^{\infty}\frac{d\omega}{2\pi}
\frac{A_{n'}(\omega')}{i\omega_{\mu}-\omega'},
\end{equation}
where the phonon spectral density is defined as
\begin{equation}
B(\bbox{k},\omega')=-2{\mathrm{Im}}D(\bbox{k},\omega').
\label{phonspectral}
\end{equation}
The frequency summation in Eq.~(\ref{mass1}) may now be performed, leading to
\begin{displaymath}
\Sigma_{n}(ip_{\nu})=-\frac{1}{N}\sum_{n',\bbox{k}}
|F_{nn'}(\bbox{k})|^{2}
\int_{-\infty}^{\infty}\frac{d\omega'}{2\pi}
\int_{-\infty}^{\infty}\frac{d\omega''}{2\pi}
[n_{\mathrm{B}}(\omega')+1]
\frac{B(\bbox{k},\omega')A_{n'}(\omega'')}{\omega'+\omega''-ip_{\nu}}.
\end{displaymath}
From the above, using Eq.~(\ref{spectral}), the retarded mass operator reads
\begin{equation}
\Sigma_{n}(\omega)=-\frac{1}{N}\sum_{n',\bbox{k}}
|F_{nn'}(\bbox{k})|^{2}
\int_{-\infty}^{\infty}\frac{d\omega'}{2\pi}
[n_{\mathrm{B}}(\omega')+1]
B(\bbox{k},\omega')G_{n'}(\omega-\omega'),
\label{mass2}
\end{equation}
and, inserting Eq.~(\ref{Gw}),
\begin{subequations}
\begin{eqnarray}\label{gamma0}
\Delta_{n}(\omega) & = &
-\frac{1}{N}\sum_{n',\bbox{k}}|F_{nn'}(\bbox{k})|^{2}
\int_{-\infty}^{\infty}\frac{d\omega'}{2\pi}
[n_{\mathrm{B}}(\omega')+1]B(\bbox{k},\omega')
\frac{\omega-\omega'-E_{n'}}{(\omega-\omega'-E_{n'})^{2}
                            +\gamma_{n}^{2}(\omega-\omega')}, \\
\gamma_{n}(\omega) & = & \frac{1}{N}\sum_{n',\bbox{k}}
|F_{nn'}(\bbox{k})|^{2}\int_{-\infty}^{\infty}\frac{d\omega'}{2\pi}
[n_{\mathrm{B}}(\omega')+1]B(\bbox{k},\omega')
\frac{\gamma_{n}(\omega-\omega')}{(\omega-\omega'-E_{n})^{2}
                            +\gamma^{2}_{n}(\omega-\omega')}.
\end{eqnarray}
\end{subequations}
To the leading order, $\Delta_{n}(\omega)$ and
$\gamma_{n}(\omega)$ may be obtained from Eq.~(\ref{gamma0},b) by
setting $\gamma_{n}(\omega)\rightarrow 0$ on the right-hand side,
which leads to
\begin{subequations}
\begin{eqnarray}
\Delta_{n}(\omega) & = &
-\frac{1}{N}\sum_{n',\bbox{k}}|F_{nn'}(\bbox{k})|^{2}
\int_{-\infty}^{\infty}\frac{d\omega'}{2\pi}
[n_{\mathrm{B}}(\omega)+1]
\frac{B(\bbox{k},\omega')}{\omega-\omega'-E_{n'}},\\
 \label{gamma}
\gamma_{n}(\omega) & = & \frac{1}{2N}\sum_{n',\bbox{k}}
|F_{nn'}(\bbox{k})|^{2}B(\bbox{k},\omega-E_{n'})
[n_{\mathrm{B}}(\omega-E_{n'})+1].
\end{eqnarray}
\end{subequations}

According to Eqs.~(\ref{phonspectral}) and (\ref{phonGreen}), the
phonon spectral density is given by
\begin{equation}
B(\bbox{k},\omega)
=\frac{8\Omega_{\bbox{k}}^{2}\,{\mathrm{Im}}\,\Pi(\bbox{k},\omega)}%
{[\omega^{2}-\Omega_{\bbox{k}}^{2}
+2\Omega_{\bbox{k}}\,{\mathrm{Re}}\,\Pi(\bbox{k},\omega)]^{2}
+[2\Omega_{\bbox{k}}\,{\mathrm{Im}}\,\Pi(\bbox{k},\omega)]^{2}}.
\label{B}
\end{equation}
The essential contribution to $\Delta_{n}(\omega)$ may be obtained
by setting $\Pi(\bbox{k},\omega)\rightarrow 0$ in Eq.~(\ref{B}); after
a simple integration one gets
\begin{displaymath}
\Delta_{n}(\omega)=\frac{1}{N}\sum_{\bbox{k},n'}
|F_{nn'}(\bbox{k})|^{2}
\left[\frac{1+N_{\bbox{k}}}{\omega-E_{n'}-\Omega_{\bbox{k}}}
+\frac{N_{\bbox{k}}}{\omega-E_{n'}+\Omega_{\bbox{k}}}\right].
\end{displaymath}
Hence, from Eq.~(\ref{Z}),
\begin{equation}\label{Zn}
    Z_{n}=1+\frac{1}{N}\sum_{\bbox{k},n'} |F_{nn'}(\bbox{k})|^{2}
\left[\frac{1+N_{\bbox{k}}}{(E_{n}-E_{n'}-\Omega_{\bbox{k}})^{2}}
+\frac{N_{\bbox{k}}}{(E_{n}-\omega-E_{n'}+\Omega_{\bbox{k}})^{2}}\right].
\end{equation}

To find $\gamma_{n}(\omega)$ we get the retarded polarization
operator $\Pi(\bbox{k},\omega)$ from Eq.~(\ref{polar2}) by
$i\omega_{\mu}\rightarrow\omega+i\delta$; thus
\begin{eqnarray}
\chi(\bbox{k},\omega) & \equiv & {\mathrm{Im}\,\Pi(\bbox{k},\omega)}
\label{chi} \\
& = & \frac{\pi}{N}\sum_{\bbox{k'}}|W(\bbox{k},\bbox{k}')|^{2}\left\{
(n_{\bbox{k}'}-N_{\bbox{k}-\bbox{k}'})\left[
\delta(\omega-\Omega_{\bbox{k}-\bbox{k}'}+\omega_{\bbox{k}'})
-\delta(\omega+\Omega_{\bbox{k}-\bbox{k}'}-\omega_{\bbox{k}'})
\right]\right. \nonumber \\
& & \left.
+(n_{\bbox{k}'}+N_{\bbox{k}-\bbox{k}'}+1)\left[
\delta(\omega-\Omega_{\bbox{k}-\bbox{k}'}-\omega_{\bbox{k}'})
-\delta(\omega+\Omega_{\bbox{k}-\bbox{k}'}+\omega_{\bbox{k}'})
\right]
\right\}. \nonumber
\end{eqnarray}
\end{widetext}

According to Eq.~(\ref{B}), the contribution from LO phonons to the
spectral density 
$B(\bbox{k},\omega-E_{n'})$ in Eq.~(\ref{gamma})
is peaked around $\omega-E_{n'}\approx\pm\Omega_{\bbox{k}}$.
Hence, the Bose function takes the values of
$n_{\mathrm{B}}(\Omega_{\bbox{k}})+1$ and
$n_{\mathrm{B}}(-\Omega_{\bbox{k}})+1=-n_{\mathrm{B}}(\Omega_{\bbox{k}})$.
If we assume that the temperature is well below 400 K, the former
is close to one and the latter is zero. Thus, $\omega\approx
E_{n'}+\Omega_{\bbox{k}}$ and we may approximate, neglecting the
LO phonon energy shift due to anharmonic coupling,
\begin{eqnarray}\label{gamma2}
\lefteqn{\gamma_{n}(\omega)=} \\
& & \frac{1}{N}\sum_{n',\bbox{k}}|F_{nn'}(\bbox{k})|^{2}
\frac{\chi(\bbox{k},\omega-E_{n'})}{[\omega-E_{n'}-\Omega_{\bbox{k}}]^{2}
+\chi^{2}(\bbox{k},\omega-E_{n'})}. \nonumber
\end{eqnarray}
There are a few effects contained in this formula. First, it generates a
broadening of the zero-phonon spectral peak at $\omega=E_{n}$ due to LO 
phonon-assisted transitions. In the absence of anharmonic coupling 
($\chi\to 0$) an energy conserving Dirac $\delta$ appears on the right-hand 
side, restricting the process to exact resonance. The anharmonicity 
releases this restriction and leads to two-phonon anharmonicity-induced
processes whenever the energy spacing matches an LO+LA or LO+TA pair, 
as described in detail in Ref.~\onlinecite{jacak02a,jacak03a}. 
Further broadening comes from the usual second-order two-phonon processes 
which are beyond the leading-order approximation used here. For the ground 
state, however, all these processes are of phonon absorption type and are
negligible at low temperatures. The second effect, which is of major 
interest to us in the present work, is related to the line shape of the
LO phonon replica  (around $\omega\approx E_{n'}+\Omega_{\bbox{k}}$) 
and its effect on the system evolution.

In this spectral region 
only the first and third terms from Eq.~ (\ref{chi}) contribute
to Eq.~(\ref{gamma2}) since the energies of LO phonons lie entirely 
above the TA branch. Out of these two, the first one vanishes at low
temperatures. Moreover, since coupling to short-wavelength modes
is favored both by the general form of the coupling ($\sim
\sqrt{k'}$, see Ref.~\onlinecite{LL10}) and by the growing density
of states, this contribution (for low $k$, as selected by the
exciton-LO-phonon form factor \cite{jacak03a,jacak03b}) is peaked
around $\omega-E_{n'}\approx 20$ meV (for GaAs, see 
Ref.~\onlinecite{strauch90} for exact dispersion curves) and gives smaller
contribution due to large denominator in Eq.~(\ref{gamma2}). The
remaining term corresponds to the LO phonon decay with emission of
another LO phonon and TA phonon. It is large, when the total
energy of the resulting phonons is close to the $\Gamma$-point LO
phonon energy, i.e. near the $L$ point in the Brillouin zone
\cite{strauch90,vallee91}. Thus, we approximate the
coupling constant and the Bose distribution by the corresponding
$L$-point values, assuming also $\bbox{k}$ near the zone center.
Replacing the summation by integration we obtain, after these
approximations,
\begin{eqnarray*}
\chi(\bbox{k},\omega) & = &
\frac{v\pi}{(2\pi)^{3}}|W(0,\bbox{k}_{L})|^{2}
(n_{\bbox{k}_{L}}+1) \\
& & \times\int d^{3}k'
\delta(\omega-\Omega_{\bbox{k}-\bbox{k}'}-\omega_{\bbox{k}'}),
\end{eqnarray*}
where $v$ is the unit cell volume. The problem is thus reduced
to finding the appropriate two-phonon density of
states.
We assume that LO and TA phonon energies may be approximated by
quadratic forms of momenta
around each L point, given by the matrices $\hat{\beta}_{\mathrm{LO}}^{(i)}$
and $\hat{\beta}_{\mathrm{TA}}^{(i)}$, respectively. The index
$i=1\ldots 4$ numbers the four pairs of opposite hexagonal walls of the
Brillouin zone (note that the four tensors differ only by orientation).
Near the $i$th $L$ point, the minimum total energy of a LO$+$TA pair with
total momentum $\bbox{k}$ is
\begin{displaymath}
\xi_{\bbox{k}}^{(i)}=\omega_{\bbox{k}_{L}}+\Omega_{\bbox{k}_{L}}
+\bbox{k} \hat{\beta}_{\mathrm{LO}}^{(i)}
(\hat{\beta}^{(i)})^{-1}
\hat{\beta}_{\mathrm{LO}}^{(i)} \bbox{k},
\end{displaymath}
where $\hat{\beta}^{(i)}
=\hat{\beta}_{\mathrm{LO}}^{(i)}+\hat{\beta}_{\mathrm{LA}}^{(i)}$.
It is clear that the LO phonon with momentum $\bbox{k}$ may decay via this
process only if
$\Omega_{\bbox{k}}>\xi_{\bbox{k}}\equiv\mathrm{min}_{i}\xi_{\bbox{k}}^{(i)}$.
One finds,
\begin{displaymath}
\int d^{3}k'
\delta(\omega-\Omega_{\bbox{k}-\bbox{k}'}-\omega_{\bbox{k}'})=
\sum_{i}\int_{S_{i}'}\frac{dS'}{2|\hat{\beta}^{(i)}\bbox{k'}|}
\end{displaymath}
where the surface $S_{i}'$ in the $\bbox{k}'$ space,
composed by all states satisfying energy and momentum conservation,
is defined by
$\xi_{\bbox{k}}^{(i)} +\bbox{k}\hat{\beta}^{(i)}\bbox{k}=\omega$
(the integral is zero if
this equation has no solutions, i.e. for
$\omega\le\xi_{\bbox{k}}^{(i)}$).
Thus, the finite bandwidth of phonon branches leads to a (lower) frequency
cutoff in the two-phonon density of final states for the anharmonic
process.

For GaAs, the tensors $\hat{\beta}^{(i)}$ may be reasonably
approximated isotropically (and hence become equal). Defining
$\beta=({\mathrm{det}}\hat{\beta}^{(i)})^{1/3}$, one finds
explicitly
\begin{displaymath}
\chi(\bbox{k},\omega) = \sum_{i}\chi^{(i)}(\bbox{k},\omega),
\end{displaymath}
where
\begin{displaymath}
    \chi^{(i)}(\bbox{k},\omega) = \frac{v}{4\pi}|W(0,\bbox{k}_{L})|^{2}
    (n_{\bbox{k}_{L}}+1) \beta^{-3/2}
\sqrt{\omega-\xi_{\bbox{k}}^{(i)}}
\end{displaymath}
for $\omega\ge\xi_{\bbox{k}}^{(i)}$ and
$\chi^{(i)}(\bbox{k},\omega) =0$ otherwise. Since, most probably,
the zone-edge LO$+$TA process is only one of many LO phonon
damping mechanisms, we shall phenomenologically add to
$\chi(\bbox{k},\omega)$ an additional contribution $\chi_{0}$,
corresponding to all the other damping processes.

As discussed above, $\gamma_{n}(\omega)$ contains various contributions, 
related to the broadened zero-phonon line, LO phonon replica, and acoustic 
phonon sidebands \cite{krummheuer02} (not included in our discussion). As 
long as these features are spectrally separated, as is the case
for the ground exciton state at low 
temperatures, they manifest themselves by processes on different timescales
($~1$ ns exponential decay \cite{borri,bayer02}, $~1$ ps overdamped 
acoustic phonon dynamics 
\cite{krummheuer02}, $~100$ fs LO phonon beats damped on $~10$ ps timescale 
as discussed below).
These processes may, therefore, be experimentally separated even if they are 
superposed on one another. As we are mostly interested in the LO phonon 
features we simply model the low-frequency sector by a Lorentzian, 
allowing for a certain broadening of the zero-phonon line. Separating this 
low-frequency part from the peak around $\omega=E_0+\Omega_{\bm{k}}$, 
we write the spectral density for the ground exciton state 
in the form
\begin{eqnarray}\label{spectral1}
A_{0}(\omega) & = & 
 2Z_{0}^{-1}\frac{\gamma_{0}}{(\omega-E_{0})^{2}+\gamma_{0}^{2}} \\
& & +\frac{2}{N}\sum_{n',\bbox{k}} \Phi_{0n'}(\bbox{k})
g_{\bbox{k}}(\omega-\Omega_{\bbox{k}}-E_{n'}), \nonumber
\end{eqnarray}
where
\begin{equation}
\Phi_{nn'}(\bbox{k})=
\frac{|F_{nn'}(\bbox{k})|^{2}}{(E_{n'}-E_{n}+\Omega_{\bbox{k}})^{2}},
\label{fi}
\end{equation}
\begin{equation}
g_{\bbox{k}}(\omega)=
\frac{\chi(\bbox{k},\omega+\Omega_{\bbox{k}})}{\omega^{2}
+\chi^{2}(\bbox{k},\omega+\Omega_{\bbox{k}})},
\end{equation}
and $\gamma_{0}$ accounts for the broadening of the zero-phonon line.
The first contribution corresponds to the central peak while the
second part describes the exciton dressing with a continuum of LO
phonon modes.

\section{The time evolution: damped LO phonon beats}
\label{sec:results}

The time-dependent correlation function for $t>0$, derived from
Eq.~(\ref{spectral1}) according to Eq.~(\ref{An}), is
\begin{eqnarray}\label{correl1}
\lefteqn{I_{0}(t)=} \\
& & Z_{0}^{-1}e^{-i(E_{0}-i\chi_0)t} +\frac{1}{N}\sum_{n',\bbox{k}}
\Phi_{0n'}(\bbox{k})
g_{\bbox{k}}(t)e^{-i(E_{n'}+\Omega_{\bbox{k}})t}, \nonumber
\end{eqnarray}
where
\begin{equation}
g_{\bbox{k}}(t)=
\int\frac{d\omega}{\pi}g_{\bbox{k}}(\omega)e^{-i\omega t}
\equiv|g_{\bbox{k}}(t)|e^{i\varphi_{\bbox{k}}(t)}.
\end{equation}
Using Eq.~(\ref{Zn}) in the low temperature limit
($N_{\bm{k}}\rightarrow 0$), one gets
\begin{widetext}
\begin{equation}
|I_{0}(t)|^{2}=\left( 1- 2\frac{1}{N}\sum_{n,\bbox{k}}
\Phi_{0n}(\bbox{k}) \right)
+ 2\frac{1}{N}\sum_{n,\bbox{k}} \Phi_{0n}(\bbox{k})
|g_{\bbox{k}}(t)|
\cos\left[ (E_{0}-E_{n}-\Omega_{\bbox{k}})t +\varphi_{\bbox{k}}(t) \right].
\label{squared}
\end{equation}
\end{widetext}
Hence, the evolution of each LO phonon mode in the polarization
cloud may be treated separately. The total effect emerges as a
superposition of such individual contributions, weighted by the
coupling strengths. It should be noted that in spite of the
growing energy separation in the denominator of the coupling
strength (\ref{fi}) (for $n=0$), the coupling to higher levels
($n'\ge 1$) is stronger than that to the $n'=0$ level because of
the charge cancellation which reduces the diagonal coupling
constants $F_{00}(\bbox{k})$ \cite{jacak03b}.

Let us study a few examples of the system evolution. The numerical results
presented below have been obtained with the GaAs parameters shown in Tab.
\ref{tab:param}. The anharmonic coupling strength
$|W(0,\bbox{k}_{\mathrm{L}})|$ was determined from the requirement that
the zone-edge LO$+$TA decay process should be responsible for 90\% of the
experimentally known decay probability of the $\Gamma$-point LO phonon.
A specific LO mode $\bbox{k}$ is damped by the zone-edge process if
$\chi(\bbox{k},\Omega_{\bbox{k}})>0$,
i.e. if its energy lies over the density of states (DOS) cutoff,
$\Omega_{\bbox{k}}>\xi_{\bbox{k}}$.
In our parabolic dispersion model one has
\begin{eqnarray*}
\lefteqn{\Omega_{\bbox{k}}-\xi_{\bbox{k}} =} \\
& &
\Omega_{0}-\Omega_{\bbox{k}_{\mathrm{L}}}-\omega_{\bbox{k}_{L}}
+\mathrm{max}_{i}\,\bbox{k}\left[
\hat{\beta}_{\Gamma}-\hat{\beta}^{(i)}_{\mathrm{LO}}
(\hat{\beta}^{(i)})^{-1} \hat{\beta}^{(i)}_{\mathrm{TA}} \right]
\bbox{k},
\end{eqnarray*}
where the last term is negative except for a very small positive
contribution (sensitive to the crudeness of dispersion modeling)
in a narrow solid angle around the $(111)$ and equivalent
directions (towards the $L$ points). Hence, if the decay process
is forbidden for the $\Gamma$-point mode it is also forbidden for
a vast majority of all the modes. Two different regimes may arise
when this process is allowed for a given mode:

\begin{figure}
\unitlength 1mm
\begin{center}
\begin{picture}(85,85)
\put(0,0){\resizebox{85mm}{!}{\includegraphics{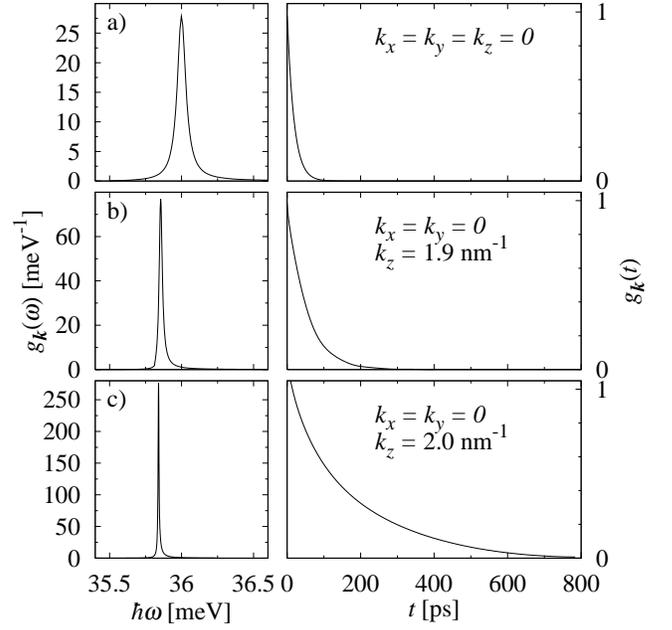}}}
\end{picture}
\caption{\label{fig:mode}Single mode line shape (left) and
temporal evolution (envelope of LO oscillations, right) for three
modes with wave vectors as shown, for
$\Omega_{0}-\Omega_{\bbox{k}_{\mathrm{L}}}-\omega_{\bbox{k}_{\mathrm{L}}}
=-0.5$ meV.}
\end{center}
\end{figure}

If the decay results in states distant from the zone edge then
the density of final states is also far from its edge. Quantitatively,
for
\begin{displaymath}
\sqrt{\Omega_{\bbox{k}}-\xi_{\bbox{k}}}\gg
\frac{v|W(0,\bbox{k}_{\mathrm{L}})|^{2}}{2\pi\beta^{3/2}}
(n_{\bbox{k}_{\mathrm{L}}}+1),
\end{displaymath}
one has
\begin{displaymath}
\frac{\partial
\chi(\bbox{k},\omega)}{\partial \omega}|_{\omega=\Omega_{\bbox{k}}}
\ll 1,
\end{displaymath}
and the value of $\chi(\bbox{k},\omega)$ is almost constant around the LO
phonon energy, leading to the Lorentzian line shape (for this specific mode)
and exponential decay (Fig \ref{fig:mode}a). The decay time of oscillations
induced by a mode in $|I_{n}(t)|^{2}$ is twice longer than the decay time
of the mode occupation (which would be governed by $|g_{\bbox{k}}|^{2}$).
In fact, the homogeneity of the ``regular'' decay (far from the DOS edge)
justifies representing all such decay processes by a single constant
$\chi_{0}$.
On the other hand, if the anharmonic decay process leads to zone-edge
states, then
$\xi_{\bbox{k}}$ is close
to $\Omega_{\bbox{k}}$ and the line shape is determined mostly by the frequency
dependence of $\chi(\bbox{k},\omega)$ and becomes non-Lorentzian (Fig.
\ref{fig:mode}b).

If the mode energy lies below the DOS cutoff, the mode
is damped only by the other processes, represented by $\chi_{0}$
(Fig. \ref{fig:mode}c). The vicinity of the DOS edge still results in
dephasing effects (by virtual processes) manifested by a
non-Lorentzian line shape.

In principle, all the three scenarios may be present among modes
forming the polaronic cloud, leading to intricate system dynamics.
However, since it is rather impossible to excite a single,
selected mode, the observable evolution always results form
superposing the dynamics of all the modes forming the polaron
cloud. Such dynamics (and the corresponding line shape) averages
the contributions from different modes. Moreover, the joint
evolution is affected by the dephasing effect resulting from
dispersion \cite{krummheuer02,jacak03b,pazy02}. Let us note that,
according to Eqs.~(\ref{spectral1},\ref{correl1}), several
contributions may be present in the system evolution, related not
only to the ground state itself but also from non-adiabatic
excitation of optically inactive excited states, the latter being much more
pronounced due to lack of charge cancellation. However, the excited
state contributions correspond to higher frequencies and may be
selectively turned off by modifying the pulse duration (i.e. its
spectral width). Below we present the results assuming the excitation of the 
ground state only and of the ground and first excited state. We neglect the 
broadening of the central line which, in the time range of interest,
would produce an additional weak slope in the time evolution.

\begin{figure}
\unitlength 1mm
\begin{center}
\begin{picture}(85,85)
\put(0,0){\resizebox{85mm}{!}{\includegraphics{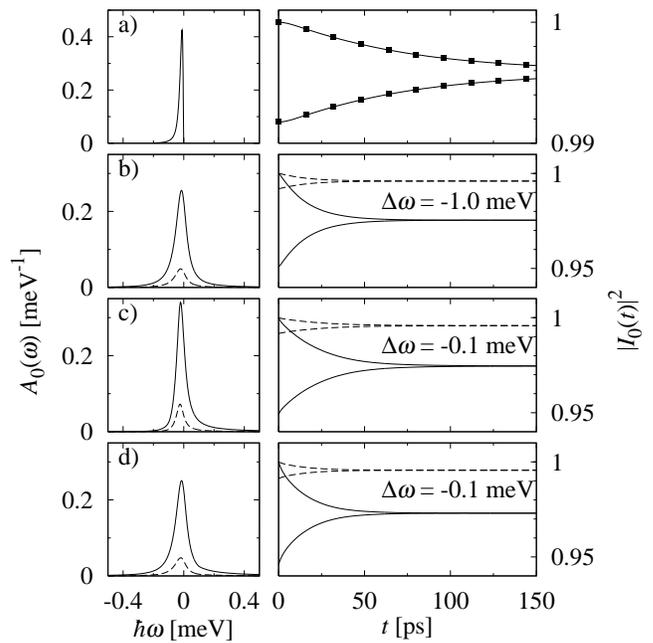}}}
\end{picture}
\caption{\label{fig:all}(a) The hypothetical line shape of the LO 
phonon replica of the ground exciton state without anharmonic damping 
(broadened only due to dispersion) and the corresponding envelope of 
LO phonon beats (line) compared to the exact result (squares). 
(b-d) The line shapes of the phonon replicas
corresponding to the ground state (dashed lines) and to the first
excited optically inactive state (solid lines) and the corresponding 
envelope of
the phonon beats for three values of the band overlap as shown.
The central frequency $\omega_{0}$ corresponds to
$E_{n}+\Omega_{0}$ for each line. In plots (b) and (c) it is
assumed that the zone-edge process contributes 90\% of the decay
probability. In (d), where the zone-edge process is forbidden, the
anharmonic LO-TA coupling is the same as in (c).}
\end{center}
\end{figure}

Before discussing the results, let us note that in the case of a
single exciton state and without any phonon anharmonicity the problem
is reduced to the independent boson model which is solvable exactly
\cite{krummheuer02}. This allows us to estimate the accuracy of our
approximate approach by comparing it with the exact result in this
limiting case. As shown in Fig.~\ref{fig:all}a 
the Green function technique gives very exact results, which is due to the 
fact that the overall phonon-induced perturbation is very weak.

The results of our calculations for the full case including anharmonic
phonon decay processes are shown in Fig.~\ref{fig:all}b-d. We calculated
the shapes of the LO replica lines for various possible energy
relations at the zone edge.
If the momentum and energy conservation for all effectively
coupled modes (with spectral width $\Delta\Omega_{\bbox{k}}$) is
satisfied by final states lying far from the zone edge, i.e. if
$\Omega_{0}-\Omega_{\bbox{k}_{\mathrm{L}}}-\omega_{\bbox{k}_{\mathrm{L}}}
\gg \Delta\Omega_{\bbox{k}}$ (large energy overlap at the zone
edge), the difference of decay times between modes becomes small
and beats corresponding to all the modes decay with the same
lifetime $\tau$, equal to the doubled lifetime of the bulk
zone-center LO phonon $\tau_{\mathrm{LO}}$. In this case, the
damping dominates over dephasing. However, since the damping and
dephasing times are still comparable, the dispersion-induced
dephasing effect is not negligible and reduces the decay rate of
the whole phonon dressing cloud oscillations below the individual
mode rate $2\tau_{\mathrm{LO}}$ (Fig. \ref{fig:all}b).

If, on the other hand, the zone-edge energy overlap
is small, then some modes will be weakly affected by the zone-edge
mechanism and some others will be unaffected by it. Thus, some modes
off the $\Gamma$ point live much longer than the zone-center mode and the
overall decay time of the phonon beats may become longer than
$2\tau_{\mathrm{LO}}$ (Fig. \ref{fig:all}c).

Finally, let us consider a situation when the LO$+$TA process is forbidden
for the zone center phonon. Although some other modes still may decay via
this process, this effect is too weak to be noticeable in the overall
dynamics and the phonon replica line again takes the Lorentzian shape with
decay time shorter than the $\tau_{\mathrm{LO}}$ due to dispersion.

\section{Conclusions}
\label{sec:conclusions}

In this paper we have studied the quantum evolution of the
exciton-LO-phonon system in an InAs/GaAs quantum dot, focusing on the
possible signature of the zone-edge LO$\rightarrow$LO$+$TA anharmonic decay
in the line shapes of the LO phonon replicas and in the time evolution of
the system.
Apart from including anharmonic damping, we have further extended the
description by incorporating contributions from optically inactive excited 
states.
We have shown that the vicinity of the edge of density of
final states leads to the appearance of long-living modes and to a
deviation from the symmetric Lorentzian line shape. In such case, the
decay time of the LO phonon beats becomes longer than
$2\tau_{\mathrm{LO}}$, where $\tau_{\mathrm{LO}}$ is  the measured decay
time of the zone-center LO phonon population. This situation appears when
the total energy of the $L$-point phonons (LO$+$TA) is only slightly lower
(by $\sim 0.1$ meV)
than the $\Gamma$-point LO phonon energy. On the other hand, when this
energy overlap becomes larger, the line shape becomes very close to
Lorentzian and the decay time of the LO phonon beats is shorter than
$2\tau_{\mathrm{LO}}$, due to additional contribution from
dispersion-induced dephasing.

In view of long coherence time of the undamped LO phonon
oscillations (due to weak dispersion), the line shapes of the
phonon  replicas and the time evolution of beats is always determined by
the anharmonic damping processes. Therefore, the damping effect
must be included in any description of carrier-LO-phonon dynamics.
Moreover, in some special cases, the detailed structure of phonon
bands may noticeably influence the measurable properties. It may
therefore be possible to extract information about the anharmonic
LO phonon damping channels from the QD spectroscopy data. Although
the phonon dispersion is obviously a material characteristics, it
may be modified by changing the material composition and
experimental conditions (e.g. pressure), so a range of dynamical
regimes might be observed.

It may be expected that an actual experiment will be performed on
the ensemble of dots by nonlinear optical techniques, e.g.
four-wave-mixing (FWM). It is known that the time-resolved FWM results
cannot be reproduced exactly by a linear spectrum or by a single
pulse dynamics \cite{vagov}. Unlike for the Markovian 
dephasing effects, in the case of the lattice response related to pure
dephasing after a fast excitation there is no simple relation between the
spectra and time-resolved evolutions obtained in linear and nonlinear 
experiments. However much similarity still remains, due
to the unique underlying physical effects. For instance, the variation of 
damping times and the non-Lorentzian shape of the LO replica reflect the 
structure of the allowed final states for an anharmonic decay process and 
may be expected to appear in any kind of experiment.
Hence, the results obtained
here are at least qualitatively relevant also for a nonlinear
experiment, although an exact extension in this direction would
certainly be desirable. Another important generalization should
consist in including final-length pulses
\cite{castella99,alicki04,machnikowski04a}.

\section*{Acknowledgments}
This paper has been supported by the Polish Ministry of Scientific
Research and Information Technology under the (solicited) grant No
PBZ-Min-008/P03/03 and by the Polish KBN under grant No. PB 2 PO3B
085 25. P.M. is grateful to the Alexander von Humboldt Foundation
for generous support.

%\bibliographystyle{prsty}
%\bibliography{abbr,quantum}

\end{document}